\documentclass[journal]{IEEEtran}
\usepackage{cite}
\usepackage{graphicx}
\usepackage{booktabs}
\usepackage{multirow}
\usepackage{array}
\usepackage{tabularx}
\usepackage{makecell}
\usepackage{adjustbox}
\usepackage{amsmath,amssymb}
\usepackage{xcolor}
\usepackage{url}
\graphicspath{{figures/}}
\DeclareGraphicsExtensions{.pdf,.png,.jpg,.jpeg}

\begin{document}

\title{Generative Modeling for Physiological Signals}

\author{Xinqi Bao, Ernest Kamavuako, and Saikat Chatterjee%
\thanks{X. Bao and S. Chatterjee are with the Department of Intelligent Systems, KTH Royal Institute of Technology, Stockholm, Sweden. E-mail: xba@kth.se.

E. Kamavuako is with the Department of Engineering, King's College London, London, UK}}

\maketitle
\begin{abstract}
Physiological signals support clinical diagnosis, health monitoring, rehabilitation, wearable sensing, and human--machine interaction. However, their applications are often constrained by limited labeled data, class imbalance, noisy or incomplete recordings, heterogeneous acquisition settings, and privacy restrictions. Generative modeling has therefore attracted increasing attention as a means of addressing some of these barriers. Recent studies have used generative models to augment scarce datasets, restore degraded recordings, translate between modalities, and synthesize conditional physiological waveforms. This review summarizes recent work on generative modeling for cardiovascular, neural, muscular, peripheral, and specialized physiological signals. Major model families are covered, including generative adversarial networks (GANs), autoencoders and variational autoencoders (AEs/VAEs), diffusion models, autoregressive sequence models, and hybrid architectures. In addition, it organizes existing evaluation practices into a hierarchical framework spanning signal-level similarity, dataset-level distribution, physiological validity, task-oriented utility, and assessments of generalization and robustness. By linking signal-specific constraints, generative roles, model families, and evaluation evidence, this review provides structured guidance for the future use and evaluation of generative models in physiological-signal research.

\end{abstract}


\section{Introduction}
Physiological signals are time-resolved measurements of bodily function acquired across diverse physiological systems. They include, for example, cardiac, neural, muscular, respiratory, autonomic and fetal systems. Depending on the physiological process of interest, these signals may be captured through different sensing principles, such as electrical, mechanical, acoustic, optical, or biochemical measurements \cite{Reddy2005BiomedicalSignalProcessing,Jeong2025Wearable}. Even within the same physiological system, different recording sites and sensing principles can reflect different aspects of function. For instance, cardiac rhythm and conduction are commonly assessed through electrocardiography (ECG), whereas acoustic recordings such as phonocardiography (PCG) capture heart sounds related to mechanical cardiac events, including valve closure \cite{Reddy2005BiomedicalSignalProcessing,Giordano2019HeartSounds, bao2020analysis}. As a result, the range of physiological-signal modalities is broad and not easily reducible to a fixed list. The widely known modalities include, for example, ECG, electroencephalography (EEG), electromyography (EMG), and photoplethysmography (PPG)\cite{Neifar2025,Jeong2025Wearable}. Because these signals preserve time-varying physiological state, they support clinical decision-making tasks such as diagnosis, monitoring, and risk stratification, as well as interactive and therapeutic applications such as rehabilitation, brain--computer interfaces, and closed-loop assistive systems \cite{Jeong2025Wearable,BCIAcquisition2024}. Their diversity, however, also creates distinct methodological requirements, since each signal family has its own morphology, sampling structure, artifact profile, missingness pattern, and clinically meaningful temporal features.

Despite their clinical and translational importance, physiological-signal datasets face recurrent constraints at both the dataset and recording levels. At the dataset level, limited sample size is a common problem, and representative data can be especially difficult to obtain for rare diseases, acute abnormal events, and clinically important edge cases \cite{Li2025FewShot}. Expert annotation is another major barrier because it is costly, time-consuming, and often dependent on specialized clinical or physiological knowledge \cite{Li2025FewShot,Sylolypavan2023Annotations}. Privacy and governance constraints further restrict data sharing and reuse \cite{Li2025FewShot,Rieke2020DigitalHealthFL}. Together, these dataset-level limitations can reduce cross-cohort, cross-site, and cross-device generalizability. At the recording level, the analytical value of physiological signals can be limited by incomplete observation and signal degradation. For example, in wearable settings, motion, poor contact, sensor displacement, and heterogeneous recording conditions can introduce artifacts, missingness, or unstable signal quality \cite{VanDerDonckt2024}. Multimodal sensing has therefore attracted increasing interest because it can provide complementary views of the same physiological state and may help contextualize, recover, or validate noisy signals \cite{John2026Multisensor}. However, it also introduces additional dependencies on temporal synchronization, cross-modal alignment, and sensor reliability \cite{John2026Multisensor,Xiao2022TimeSynchronization}. Together, dataset- and recording-level limitations can weaken model performance, transferability, and generalization.

Generative modeling offers one response to these constraints by learning empirical signal distributions and using them to produce new samples, conditional reconstructions, or missing components. In machine learning, this direction has developed rapidly through several families of models, including generative adversarial networks (GANs), variational autoencoders (VAEs), diffusion-based models, and autoregressive sequence models \cite{Goodfellow2014GAN,Kingma2014VAE,Ho2020DDPM,VanDenOord2016WaveNet}. These model families have increasingly been adapted to physiological-signal research, where generation is used not only for de novo waveform synthesis but also for augmentation, restoration, cross-modal generation, and controlled synthesis \cite{Neifar2025,He2025TimeVarying,Loni2025Review}. Thus, generation in the current field is better understood as a set of distributional, reconstructive, translational, and conditional modeling tasks rather than as waveform synthesis alone.

These roles differ in what generated data must preserve and in the evidence needed to support the intended claim. Augmentation aims to broaden training variation and should preserve label-relevant physiological information without introducing artificial cues \cite{Berger2023,Zanchi2025SyntheticECG}. Restoration should recover missing or corrupted content without removing subject-specific structure, abnormal morphology, or transient events. Cross-modal generation should produce a target signal that is plausible in itself and physiologically consistent with the source. Controlled synthesis makes a stronger claim: the generated signal should express a specified clinical or physiological condition while remaining credible for its intended use \cite{bao2026diffusion}. Augmentation, restoration, and cross-modal generation are therefore commonly functional uses, in which generation supports another task, whereas physiological synthesis places the main claim on the generated signal itself. This distinction explains why different roles require different combinations of similarity metrics, physiological checks, downstream evaluation, and expert assessment.

Generative physiological-signal research has expanded rapidly in recent years \cite{Neifar2025,He2025TimeVarying,Loni2025Review}, but studies remain difficult to compare because they differ in signal type, recording context, generative objective, model family, and evaluation protocol. A model-centered or single-modality summary cannot capture this variation: the same architecture may serve several functional roles, while the same role may impose different physiological constraints across signals. This review therefore links four dimensions across a broad range of physiological signals: signal family, functional role, model family, and evaluation evidence. A structured corpus maps broad patterns in modality coverage, generative roles, model families, and publication trends, while representative studies are examined in depth to show how these dimensions interact. By connecting intended use with signal-specific constraints, model design, and evaluation evidence, the review aims to provide a clearer and more practical account of generative modeling in physiological-signal research.

\section{Scope, Corpus, and Conceptual Organization}

This is a structured review combining corpus-based mapping with representative in-depth synthesis, rather than a formal systematic review or meta-analysis. To support both field-level mapping and detailed synthesis, this review used a two-level corpus design. A keyword-based search of Web of Science, Scopus, PubMed, IEEE Xplore, and arXiv identified 379 records published between January 2015 and December 2025. After deduplication and screening of titles, abstracts, and available full texts, records were excluded if they did not involve physiological waveform or time-series data, lacked a substantive generative-modeling component, focused only on imaging or text, or mentioned synthetic data only peripherally. This process retained 256 studies as the content-eligible corpus for summarizing broad patterns in signal modality, functional role, model family, and publication trend. A representative subset of 120 studies was selected for in-depth analysis. The selection was designed to preserve coverage across major signal families, generative roles, and model families while reducing redundancy among studies addressing similar problems. The remaining content-eligible studies were retained for corpus-level mapping and trend analysis. The search strategy and corpus construction are described in Appendix~\ref{app:search}.

The review is organized around four dimensions. First, physiological signals are grouped pragmatically into cardiovascular, neural, muscular, peripheral, and other specialized signals. Second, generative roles are grouped into augmentation, restoration, cross-modal generation, and synthesis. Third, model families are summarized by dominant design patterns, including GAN-based, AE/VAE-based, diffusion-based, autoregressive or sequence-based, and hybrid approaches. Fourth, evaluation evidence is considered in terms of signal similarity, distributional realism, physiological plausibility, downstream utility, and generalization or robustness. Given the diversity of physiological signals and the heterogeneity of generative-model architectures, these dimensions are intended as pragmatic organizing categories rather than strict taxonomies. They are used to improve comparability and readability across a broad and uneven literature.

\begin{figure*}[!t]
\centering
\includegraphics[width=\textwidth]{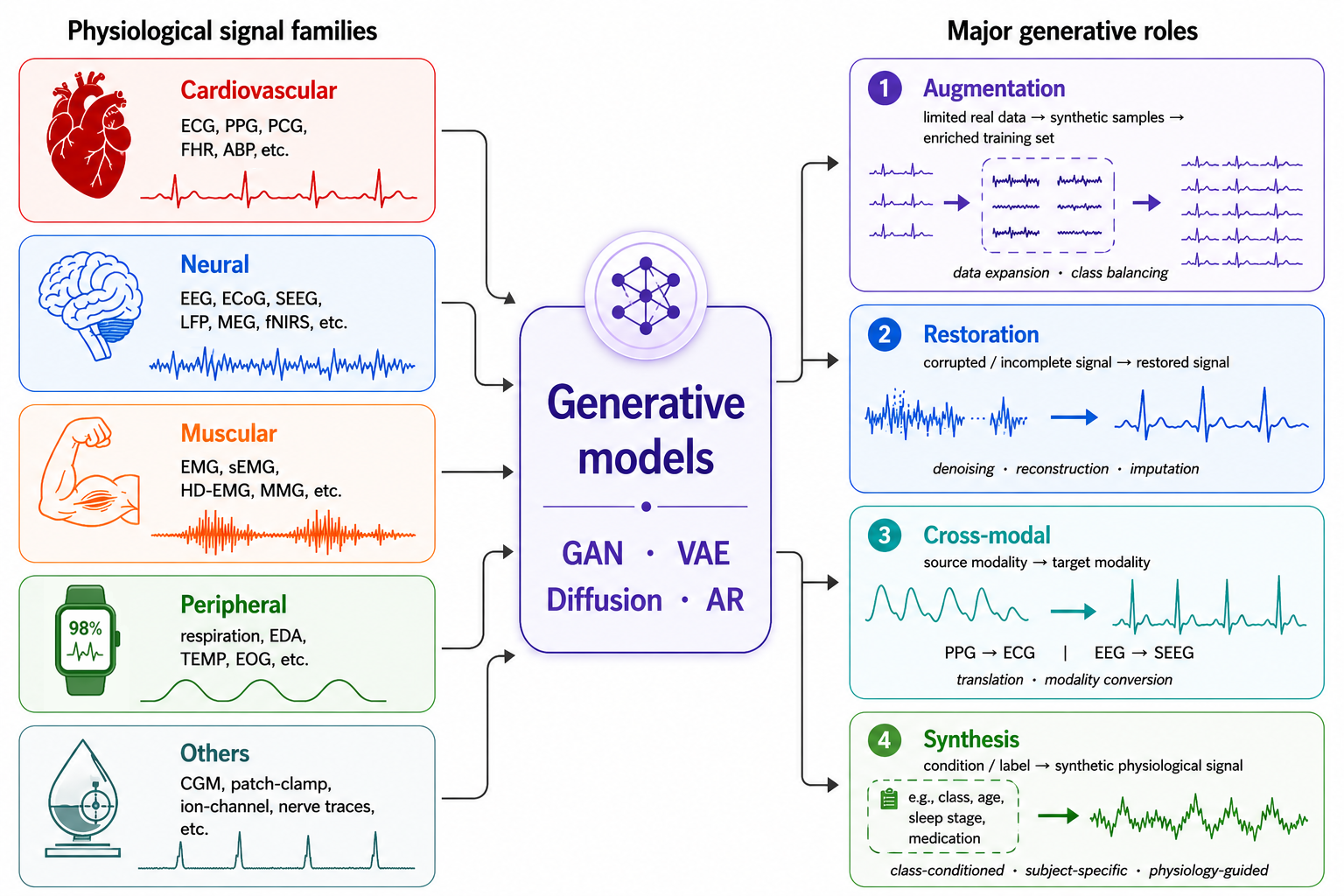}
\caption{Conceptual organization of generative modeling for physiological signals.
Signals are grouped into cardiovascular, neural, muscular, peripheral, and other specialized families, and generative models are organized by four major functional roles: augmentation, restoration, cross-modal generation, and synthesis.
These signal families are pragmatic rather than strict biological taxonomies.
Peripheral and other specialized signals are retained because the current literature is sparse but not absent, and because they represent emerging applications of generative physiological-signal modeling.}
\label{fig:conceptual}
\end{figure*}

\begin{figure}[!t]
\centering
\includegraphics[width=\columnwidth]{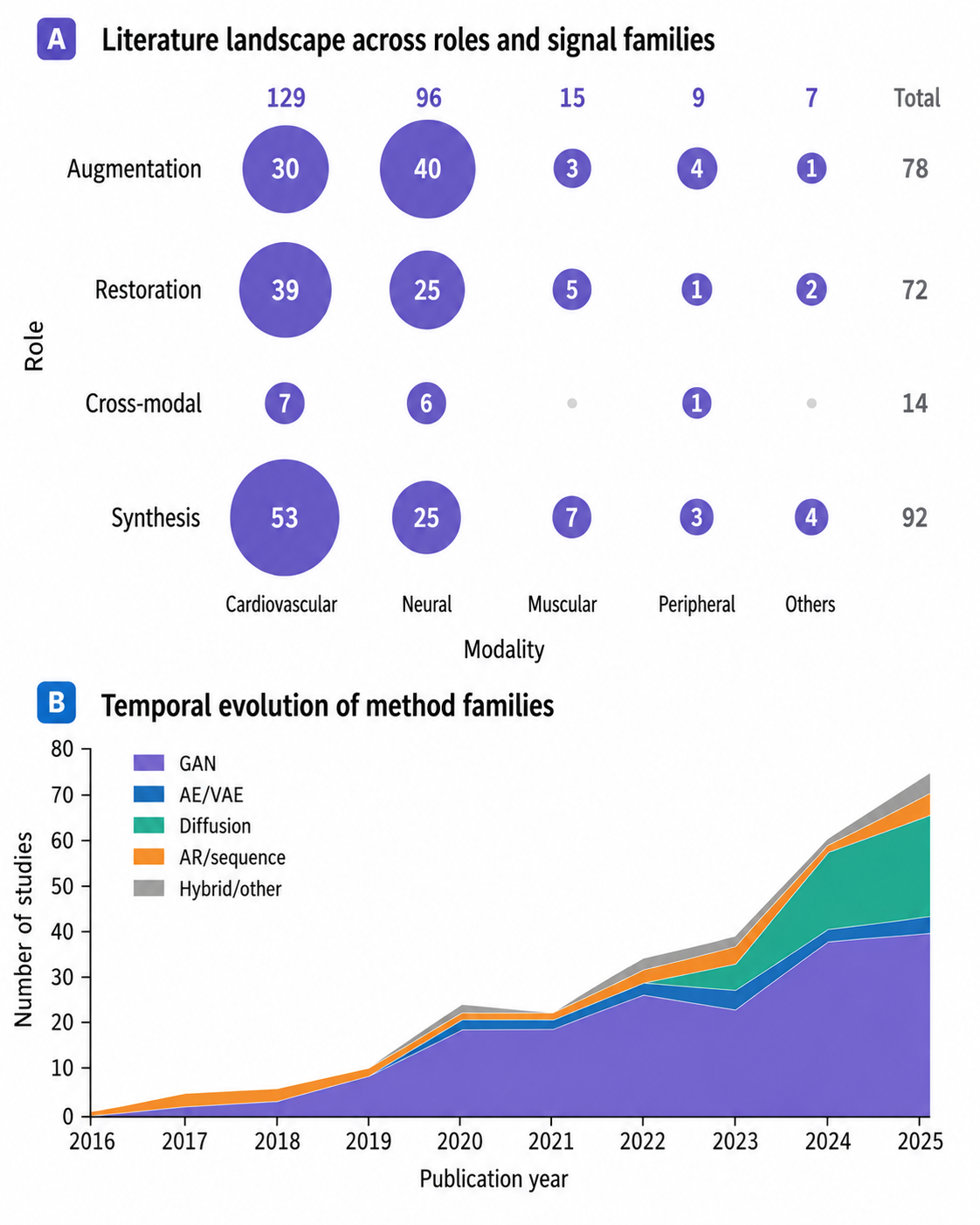}
\caption{Literature landscape of the content-eligible corpus. (A) Role--modality distribution of studies. Bubble size and labels indicate study counts. Column and row totals summarize signal-family and role totals. (B) Temporal evolution of model families in the corpus.}
\label{fig:landscape}
\end{figure}

Figure~\ref{fig:conceptual} summarizes this organization.
Figure~\ref{fig:landscape} provides a corpus-level overview of the 256 content-eligible studies.
The role--modality matrix shows that cardiovascular and neural signals account for much of the literature, while muscular, peripheral, and specialized signals remain less represented.
The temporal summary shows the continued prominence of GAN-based methods and the more recent emergence of diffusion-based and hybrid approaches.
These patterns are used as a map for the remainder of the review rather than as an exhaustive ranking of model quality.

\section{Signal Characteristics and Generative Modeling Constraints}
\subsection{Cardiovascular waveforms}
As shown in Fig.~\ref{fig:landscape}, cardiovascular signals account for a large share of generative physiological-signal studies. The cardiovascular family includes signals that reflect different physiological processes, such as cardiac electrical activity, peripheral pulse-wave propagation, acoustic mechanical events, and pressure-wave dynamics. Commonly studied examples include ECG, PPG, PCG, heart-rate, and arterial blood pressure (ABP). These signals are acquired through different sensing approaches, support different physiological analyses, and face different data or acquisition bottlenecks.

ECG records cardiac electrical activity and is widely used to assess rhythm disorders, conduction abnormalities, ischemic changes, and other clinically relevant cardiac states \cite{dong2025domain,bao2022paroxysmal}. Its clinical interpretation depends on rhythm and interval structure, as well as the morphology of the P wave, QRS complex, ST segment, and T wave \cite{Berger2023,CardioGAN2021}. ECG research benefits from several widely used public databases, such as MIT-BIH Arrhythmia, PTB-XL, and SPH \cite{Moody2001MITBIH,Wagner2020PTBXL,liu2022large}. However, clinically important abnormal patterns are often much less frequent than normal classes. For example, PTB-XL contains 21,799 ECG recordings, but only 73 recordings annotated with atrial flutter and 27 with supraventricular tachycardia \cite{Wagner2020PTBXL}. This type of imbalance can limit supervised learning and motivates the use of generative models to expand underrepresented patterns. At the same time, generated ECGs can be misleading if they appear rhythmic while distorting intervals, beat morphology, or abnormal patterns that are critical to clinical interpretation. In ambulatory or long-term ECG monitoring, motion artifacts, poor electrode contact, and missing segments further create restoration needs, where generative models may be used to recover degraded or incomplete recordings.

PPG reflects peripheral blood-volume changes and pulse-wave propagation \cite{Allen2007PPG,Park2022PPG}. It is attractive for wearable and continuous monitoring because it is easy to acquire and can support pulse-rate monitoring, vascular assessment, and blood-pressure-related estimation \cite{Castaneda2018WearablePPG,Charlton2022WearablePPG,Kim2023PPGWearables}. Its relevant structures include pulse contour, peak timing, inter-pulse intervals, amplitude variation, and beat-to-beat consistency. However, PPG is strongly affected by measurement site, contact pressure, vascular state, motion, and device conditions \cite{Allen2007PPG,Scardulla2023PPGLimitations,Pham2025PoorSkinContact}. Compared with ECG, PPG studies are often more dependent on specific acquisition protocols, which can limit cross-study comparability and generalization \cite{Scardulla2023PPGLimitations,Charlton2022PPGBestPractices}. Generative models are therefore particularly relevant for restoring noisy recordings and supporting cross-modal estimation from PPG to accessible cardiovascular signals.

Heart-rate trajectories represent a different form of cardiovascular signal because they summarize cardiac timing over longer windows rather than preserving detailed beat morphology.
Fetal heart rate within cardiotocography (CTG) is a representative example. It is used for fetal monitoring, and its interpretation relies on temporal patterns such as baseline, variability, accelerations, and decelerations, together with their relation to associated signals such as uterine activity \cite{FHRGAN2022,AyresDeCampos2015FIGO}. These recordings may be affected by signal loss, maternal--fetal signal confusion, and recording artifacts. For heart-rate trajectories, useful generation should therefore preserve clinically meaningful temporal organization and state-dependent patterns over longer windows, rather than only producing locally smooth or periodic sequences.

PCG captures acoustic events related to mechanical cardiac function.
It is used for structural assessment, where the first and second heart sounds (S1 and S2), systolic--diastolic ratio, and murmur patterns are important \cite{Homaeinezhad2012,Giordano2019HeartSounds,bao2022effect,xu2022hierarchical,bao2023time}. Compared with ECG and PPG, PCG contains higher-frequency acoustic content and is more sensitive to recording conditions. PCG recordings are normally affected by ambient noise, respiratory sounds, friction, sensor contact, and device characteristics \cite{bao2024signal}. Although public heart-sound resources have grown, their scale, annotation conventions, and population coverage remain less standardized than major ECG resources. Public datasets such as the 2016 PhysioNet/CinC Challenge heart-sound data and the CirCor DigiScope dataset have improved access to PCG recordings, but murmur and disease-specific categories still represent only a subset of available recordings \cite{PhysioNetCinC2016HeartSound,CirCor2022DigiScope}. More detailed pathological categories, such as specific valve lesions, are rarer still.
This scarcity can limit the generalizability of supervised models trained for heart-sound classification. This makes generative modeling relevant for both data augmentation and restoration, but it also creates a specific risk. A model that removes noise while suppressing murmur-related structure may produce a cleaner waveform that is less clinically useful. For PCG, therefore, the key constraint is preservation of acoustic events and pathological sound patterns under noisy recording conditions \cite{bao2026diffusion}.

Taken together, these examples show that cardiovascular signals face multiple forms of generative-modeling constraints. ECG, PPG, heart-rate trajectories, and PCG each encode different layers of cardiovascular physiology and expose different limitations in data availability, class distribution, acquisition quality, or temporal coverage. Generative models may help address these limitations, but their outputs are useful only when they preserve the signal-specific structures that carry physiological or clinical meaning.

\subsection{Neural signals}
Neural signals account for another large share of generative physiological-signal studies.
Unlike cardiovascular waveforms, which often involve more explicitly periodic structures and more diverse physiological targets, neural recordings mainly aim to capture the spatiotemporal dynamics of brain or neural activity. Their measurement settings differ in spatial scale, depth, and invasiveness, but their meaningful information is commonly expressed through time-varying spectral content, transient events, spatial or cross-channel dependencies, and state-related dynamics \cite{You2025EEGReview,LopesDaSilva2013}.

These constraints become clearer when neural recordings are viewed as a continuum of measurement settings. Scalp EEG is non-invasive and widely used for sleep staging, seizure analysis, cognitive-state monitoring, and brain--computer interfaces \cite{LopesDaSilva2013,Zhang2023WearableBCI}. However, scalp EEG has limited source localization because recorded potentials are spatially mixed by volume conduction and are affected by electrode montage, impedance, and non-neural artifacts \cite{LopesDaSilva2013,Burle2015EEGResolution,Seok2021Artifacts}. Electrocorticography (ECoG) and stereoelectroencephalography (SEEG) are invasive intracranial recordings that provide more localized information, with ECoG measuring activity from the cortical surface and SEEG sampling deeper brain structures or distributed epileptic networks \cite{Parvizi2018IntracranialEEG,E2SGAN2022}. This improved localization is valuable for epilepsy mapping and functional localization, but electrode layouts are sparse, patient-specific, and determined by clinical needs rather than standardized whole-brain sampling. Local field potentials (LFPs) provide still more localized recordings from implanted electrodes, but they are usually available only in small, task-specific, or disease-specific cohorts. Beyond electrical recordings, magnetoencephalography (MEG) non-invasively measures magnetic fields generated by neural activity and provides complementary spatiotemporal information, but it requires specialized instrumentation and is less common in large open datasets \cite{Baillet2017MEG}. Functional near-infrared spectroscopy (fNIRS) measures hemodynamic responses associated with neural activity. It is portable and suitable for functional monitoring, but has lower temporal resolution and is sensitive to systemic physiology, motion, and optode contact \cite{Yucel2021FNIRSBestPractices,Klein2024FNIRSSpecificity}. Across these measurement settings, data limitations arise from small cohorts, subject- and session-specific variability, non-standard channel layouts, missing channels, and differences in recording hardware or experimental protocols.

These characteristics make generative modeling relevant to neural signals, but they also make evaluation difficult. Generative models may enrich limited datasets, remove artifacts, reconstruct missing channels, generate virtual channels, or translate between source and target neural domains \cite{Neifar2025,E2SGAN2022,Svantesson2021}. The risk is that a generated signal may look smooth or statistically similar in the time domain while losing the neural information that matters. For example, artifact removal can suppress frequency-band activity or event-related components, channel reconstruction can break cross-channel dependence, and cross-domain generation can produce plausible traces that do not preserve target-domain spatial organization. This problem is amplified by nonstationarity across subjects, sessions, cognitive states, tasks, and recording conditions, as well as by ocular, muscle, motion, line-noise, and wearable artifacts \cite{Seok2021Artifacts}. For neural signals, the key generative-modeling constraint is therefore preservation of spectral--spatial--state structure under nonstationarity, artifact contamination, and heterogeneous channel coverage \cite{LopesDaSilva2013,Li2025EEGPlausibility}.

\subsection{Muscular signals}
Muscular signals, especially electromyography (EMG), form a more application-driven setting. Surface EMG records electrical activity associated with muscle activation through electrodes placed on the skin, whereas intramuscular EMG provides a more localized but invasive view of muscle activity \cite{Campanini2020,Kamavuako2014}. Unlike cardiovascular waveforms, EMG is not a regular periodic template. Compared with neural signals, it is often more directly linked to peripheral motor execution, although it remains strongly affected by recording conditions. EMG is typically expressed as burst-like, task-dependent activity whose amplitude, frequency content, timing, and spatial distribution depend on movement type, contraction intensity, fatigue, electrode placement, and subject-specific physiology \cite{Campanini2020,Betthauser2020}.

EMG is widely used in gesture recognition, prosthetic control, rehabilitation assessment, human--machine interfaces, and robotic or orthotic control \cite{Kamavuako2014,Betthauser2020,ChatEMG2025}. In these settings, the signal is useful because it preserves information about motor intent and muscle activation. However, the mapping between EMG and movement can change substantially across subjects, sessions, and recording conditions. Skin impedance, electrode shift, skin moisture, subcutaneous fat, muscle fatigue, and inter-muscle crosstalk can all alter the recorded signal and reduce cross-session or cross-subject classification performance \cite{Betthauser2020,Lee2024DailyEMG,Chowdhury2013EMG}. As a result, practical EMG systems often require repeated calibration, multi-session training, or adaptation to new users and recording conditions.

Generative models are therefore relevant for improving session or subject variability, restoring noisy recordings, and supporting adaptation across recording conditions \cite{Coelho2023,Chen2022EMG,ChatEMG2025}. However, these uses should be evaluated by functional realism rather than local waveform realism alone. For muscular signals, a generated signal is useful only if it remains consistent with the motor intent, movement class, or control target that the downstream system is expected to infer.

\subsection{Peripheral signals}
In this review, peripheral signals are treated as a pragmatic, literature-driven category rather than as a homogeneous physiological family. The term refers to non-cardiac, non-neural, and non-muscular physiological streams that are commonly acquired through peripheral or wearable sensing. Examples in the current generative literature include respiratory signals, electrodermal activity (EDA), skin temperature (TEMP), electrooculography (EOG), and related wearable physiological streams. These signals differ in physiological origin and waveform structure. They are discussed together because the current generative literature on them is smaller and more task-specific than the better-established ECG, EEG, EMG, and PPG literatures. Several representative studies have nevertheless begun to explore generative modeling for respiratory augmentation, wearable stress-data synthesis, privacy-preserving sensor generation, and EOG generation \cite{Jayalakshmy2021RespiratoryCGAN,Aqajari2021PPGRespirationCycleGAN,Saldanha2022RespiratoryVAE,Ehrhart2022WearableStressCGAN,Lange2024WearablePrivacy,Choi2025EOGDiffusion}.

Respiratory signals are the most physiologically structured examples in this group. Respiration is a central physiological process that supports gas exchange together with cardiovascular circulation, and respiratory rate is a widely used vital sign in clinical monitoring and deterioration assessment \cite{Cretikos2008RespiratoryRate}.
Respiratory activity can be measured directly using airflow sensors, respiratory belts, impedance-based methods, or acoustic recordings, and it can also be derived indirectly from related physiological signals such as ECG or PPG \cite{Vitazkova2024RespiratoryMonitoring,bao2020comparison,bao2020estimation}. Compared with many cardiac waveforms, respiration is quasi-periodic but more strongly modulated by behavioral state, voluntary control, physical conditions. Relevant structures include respiratory rhythm, cycle morphology, amplitude variation, pauses, irregular breathing patterns, and disease-related acoustic characteristics. Generative models have been used to augment respiratory signals or respiratory sounds for classification, and to reconstruct respiratory information from related wearable streams such as PPG \cite{Jayalakshmy2021RespiratoryCGAN,Aqajari2021PPGRespirationCycleGAN,Saldanha2022RespiratoryVAE}. In these settings, useful generation should preserve cycle-level respiratory organization or disease-relevant acoustic features.

Wearable autonomic and affective streams introduce a different set of constraints. EDA and TEMP are commonly used in stress, arousal, and affective-state monitoring, often together with other wearable sensor channels \cite{Kyriakou2019WearableStress,Almadhor2023EDAStress}.
These signals are usually lower-dimensional and slower-varying than ECG or EMG, but they are strongly influenced by subject, activity, device, skin contact, environmental conditions, and recording protocol. This makes cross-subject or cross-study generalization difficult and makes augmentation or robustness testing relevant. At the same time, wearable stress data can be difficult to share because they may encode sensitive behavioral and physiological information, which motivates privacy-preserving synthetic sensor data. Generative models have therefore been explored for wearable stress-data augmentation and privacy-aware synthesis \cite{Ehrhart2022WearableStressCGAN,Lange2024WearablePrivacy}. In these applications, evaluation should not only test whether generated signals look plausible, but also whether they preserve stress- or affect-relevant trends, cross-sensor relationships, and, when data sharing is claimed, an acceptable utility--privacy trade-off.

EOG provides a further peripheral example in which generation is linked to discrete event structure. Because EOG reflects eye-movement-related bioelectric activity, generated EOG should preserve the timing, direction, and class-specific patterns of eye movements that support downstream classification tasks \cite{Choi2025EOGDiffusion}.

Peripheral-signal studies rely on protocol-specific datasets, limited participants, and heterogeneous sensors rather than large standardized databases. Generative models may therefore be useful for expanding scarce task-specific data, testing robustness to acquisition variability, reconstructing missing or indirect signals, and enabling privacy-aware sharing.

\subsection{Other specialized physiological signals}

Other specialized physiological signals are retained as a separate pragmatic category for physiological time series and experimental traces that do not fit naturally into the cardiovascular, neural, muscular, or peripheral groups. Representative examples include continuous glucose monitoring (CGM), patch-clamp or ion-channel recordings, and nerve traces. These signals are not grouped by a shared waveform morphology. Instead, they are grouped because generative modeling is often motivated by specialized data constraints, such as strong subject or protocol dependence, rare events, limited data access, and costly annotation.

CGM time series provide the clearest metabolic example. CGM records interstitial glucose dynamics and is used in diabetes management, glucose forecasting, hypoglycemia-risk assessment, and personalized simulation. Unlike short physiological waveforms, glucose trajectories are shaped by meals, insulin, physical activity, circadian patterns, stress, illness, and individual metabolic response \cite{Kim2023CGMPhysicalActivity,Phillips2023PersonalizedGlucose}. This creates a strongly personalized and event-driven time series, where clinically important episodes such as hypoglycemia may be underrepresented. Generative models have therefore been used to synthesize continuous glucose profiles, generate personalized glucose trajectories, produce conditional blood-glucose profiles, and augment data for hypoglycemia prediction \cite{Cichosz2022CGMCGAN,Zhu2023GluGAN,Mujahid2022GlucoseCGAN,Seo2024CGMAugmentation}.
For these applications, generated data should preserve temporal glucose dynamics, subject-specific patterns, clinically plausible ranges, and event timing.

Patch-clamp and ion-channel recordings represent a different specialized setting. These traces capture cellular or single-channel electrophysiological activity under controlled experimental conditions. Their analytical value often depends on event transitions, open--closed state structure, dwell times, and high-quality annotation. Because manual labeling and idealization can be labor-intensive, generative models can be useful when they produce annotated traces for detector training or method validation \cite{Ball2022DeepGANnel}.
In this case, generated data are valuable not simply because they resemble real traces, but because they preserve annotation-relevant structures that support downstream analysis.

Nerve recordings provide a further specialized example, although the current evidence base is still limited. Spontaneous vagus nerve recordings, for instance, can be low-amplitude, noisy, and difficult to decode \cite{Ribeiro2023VagusNerveML}. Machine-learning approaches have therefore been explored for denoising and decoding such recordings under challenging signal-quality conditions \cite{Ribeiro2023VagusNerveML}. This example is closer to restoration and representation learning than to mature synthetic waveform generation, but it still illustrates a broader constraint in specialized physiological traces: useful modeling often depends on making weak, noisy, or difficult-to-label signals analytically usable.

Overall, other specialized physiological signals show a different motivation for generative modeling than the major waveform families. The goal is often not to build a general-purpose signal generator, but to address personalization, protocol dependence, annotation burden, rare-event scarcity, or limited data access. Evaluation should therefore be tied to downstream analytical value, such as glucose-event prediction, subject-specific trajectory realism, detector training, annotation quality, or denoising utility.

\section{Functional Roles of Generative Modeling in Physiological Signal Research}

\subsection{Augmentation}

Augmentation is one of the most direct uses of generative modeling when real datasets provide insufficient coverage for robust learning. This situation is common when abnormal events, rare disease classes, subject-specific patterns, or movement classes are underrepresented. In this role, synthetic signals are introduced into the training process to broaden the observed data distribution, improve class balance, or expose a downstream model to additional task-relevant variation. The generated signal is usually not the final scientific object. Its value lies in whether it improves learning under limited-data or imbalanced-data conditions without introducing artificial cues or leakage.

Electrocardiography provides a clear example. Abnormal cardiac rhythms and pathological patterns can be much less frequent than normal recordings, which motivates synthetic generation for minority-class enrichment and imbalanced ECG classification \cite{Yang2024CECG,Tang2025CCGAN}. More general ECG synthesis studies have also evaluated whether generated beats can support data augmentation for downstream classification \cite{Adib2025SyntheticECG}. Similar logic appears in sleep EEG. SleepGAN generated both EEG epochs and stage-transition sequences under few-shot conditions, and the augmented data improved sleep-staging performance compared with using limited real data alone \cite{SleepGAN2022}. In surface EMG, WGAN-GP- and DCGAN-based studies generated synthetic movement or hand-motion samples to enlarge the training set and improve downstream classification \cite{Coelho2023,Chen2022EMG}. ChatEMG extends this logic into a control-oriented setting by generating condition-, session-, or subject-specific EMG signals to improve intent inference for robotic hand-orthosis control \cite{ChatEMG2025}. Less represented signals follow the same rationale: generative models have been used to augment respiratory signals or respiratory sounds, wearable stress-sensor data, and CGM-based hypoglycemia prediction data \cite{Jayalakshmy2021RespiratoryCGAN,Saldanha2022RespiratoryVAE,Ehrhart2022WearableStressCGAN,Seo2024CGMAugmentation}.

These examples show that augmentation is not simply making more data. It is an attempt to improve coverage of physiologically or functionally meaningful variation. For ECG, this may mean rare arrhythmia morphology; for EEG, underrepresented sleep or cognitive states; for EMG, movement classes or new subject/session conditions; and for wearable or metabolic signals, task-specific states or rare events. The main risk is that synthetic samples may improve a classifier through spurious dataset-specific cues, leakage, or oversampling of narrow patterns rather than by adding meaningful physiological diversity. Leakage can occur when near-duplicate segments cross train--test boundaries, when synthetic training samples are derived from subjects or recordings that also appear in testing, or when preprocessing and model selection use information from the evaluation set. This risk is particularly important when augmentation is evaluated only on internal splits derived from the same source dataset. Augmentation therefore requires evaluation beyond downstream accuracy alone, including checks for diversity, class-conditional validity, subject- or session-independent splits, and, where possible, robustness on external data.

\subsection{Restoration}

Restoration addresses a different weakness: recordings may exist, but their quality, continuity, or structural completeness may be insufficient for reliable analysis. This role includes denoising, artifact removal, missing-segment recovery, imputation, channel reconstruction, and related completion tasks. Unlike augmentation, restoration does not primarily seek to enlarge the dataset. It uses generation as predictive reconstruction, aiming to recover usable signal content from degraded, noisy, or incomplete observations.

This role appears across several physiological settings. In fetal or high-noise ECG, adversarial denoising has been proposed to suppress noise while preserving clinically relevant waveform structure \cite{Mvuh2024Denoising}. In EEG, channel reconstruction and virtual-channel generation have been used to compensate for limited-channel acquisition or missing electrodes, with the goal of recovering part of the lost spatial information \cite{Perez2025EEGChannel,Svantesson2021}. In wearable physiological monitoring, missing data and non-wear periods motivate generative imputation or reconstruction of heart-rate and other sensor streams \cite{VanDerDonckt2024,Lin2020WearableMissing}. Restoration also appears in EMG, where noise, artifacts, and interference can obscure task-relevant muscle activation \cite{Boyer2023EMGNoise,Chowdhury2013EMG}. Specialized traces provide a related example: low-amplitude nerve recordings may require denoising or decoding before they become analytically usable \cite{Ribeiro2023VagusNerveML}.

The key point is that restoration is not equivalent to generic signal smoothing. A restored physiological signal must preserve the structures that matter for subsequent interpretation.
For ECG, this may involve beat morphology, interval structure, and abnormal waveform events.
For EEG, it may involve spectral content, transient events, and channel relationships.
For EMG, it may involve burst timing and movement-related activation.
For wearable or specialized traces, it may involve temporal continuity, event timing, or weak activity that is easily suppressed as noise.
A low reconstruction error is therefore not sufficient if the model removes clinically or functionally meaningful events or replaces uncertain signal content with an overly smooth continuation. Restoration-oriented evaluation should combine signal-level error or similarity metrics, such as root mean square error (RMSE), signal-to-noise ratio (SNR), percentage root-mean-square difference (PRD), and correlation, with checks on morphology, spectral content, event preservation, uncertainty, and downstream diagnostic or task performance.

\subsection{Cross-modal generation}

Cross-modal generation refers to generating one physiological modality from another. This role is motivated by the fact that some signals are easier, cheaper, less invasive, or more wearable than others, while the target signal may contain richer physiological or clinical information. Unlike restoration, the target is not a degraded version of the same signal, but a related and non-equivalent physiological measurement. The model is therefore learning a cross-signal mapping rather than simply reconstructing a missing trace.

PPG-to-ECG translation is a representative example. PPG is easier to collect continuously in wearable settings, whereas ECG contains more direct cardiac electrical information. CardioGAN framed this problem as generating ECG waveforms from PPG input, and later diffusion-based work extended the same direction with higher-fidelity PPG-to-ECG translation \cite{CardioGAN2021,RDDM2024}. A related peripheral example is PPG-to-respiration generation, where CycleGAN-based modeling has been used to reconstruct respiratory signals from raw PPG for respiratory-rate estimation \cite{Aqajari2021PPGRespirationCycleGAN}.
Cross-modal generation also appears in neural recordings. E2SGAN generated SEEG from simultaneous EEG data, addressing a setting where invasive target recordings are more informative but much harder to acquire \cite{E2SGAN2022}. More general cross-modal self-supervised learning methods, such as CroSSL, further show how missing modalities can be handled by learning relationships across physiological streams \cite{CroSSL2024}.

The main challenge in cross-modal generation is that physiologically related signals are not interchangeable. A generated ECG from PPG, for example, should not only resemble an ECG waveform but also preserve cardiac information that is recoverable from the source PPG. Similarly, EEG-to-SEEG translation should retain clinically or functionally relevant target-domain structure without hallucinating features unsupported by the input signal. This information asymmetry is important because the target modality may contain details that cannot be inferred from the source alone. Cross-modal generation therefore requires both target-signal fidelity and source--target consistency; waveform similarity alone may not show whether the generated signal reflects information actually present in the source.

\subsection{Synthesis}

Synthesis refers to de novo or condition-controlled generation of physiological signals. In this role, the synthetic signal is treated as an object for simulation, scenario construction, benchmarking, controlled exploration, or data sharing. Synthesis overlaps technically with augmentation, restoration, and cross-modal generation because all of these roles produce synthetic signal content. However, the emphasis is different. In synthesis, the central question is whether the model can produce realistic, diverse, controllable, and physiologically meaningful signals that can be interpreted or used as standalone synthetic physiological objects.

ECG synthesis has been widely studied, as summarized in recent scoping work \cite{Zanchi2025}. Some studies focus on general ECG-like waveform generation \cite{Adib2025SyntheticECG}, whereas others support conditional or rhythm-abnormality-aware ECG synthesis \cite{ECGAN2023}. More recent work has moved toward controllable synthesis.
TransDiffECG, for example, was designed for semantically controllable ECG generation, allowing customized physiological details rather than unconstrained sampling alone \cite{TransDiffECG2025}. Fetal monitoring provides another example. FHRGAN generated fetal heart-rate sequences of different physiological and pathological categories and arbitrary lengths in low-resource settings \cite{FHRGAN2022}. Metabolic time series provide a specialized synthesis setting, where GAN-based models have been used to generate continuous glucose profiles, personalized glucose trajectories, and conditional blood-glucose sequences for diabetes-related modeling \cite{Cichosz2022CGMCGAN,Zhu2023GluGAN,Mujahid2022GlucoseCGAN}. In experimental electrophysiology, DeepGANnel generated fully annotated patch-clamp traces from small labeled seed data, supporting detector training where manual sample-by-sample annotation is difficult to scale \cite{Ball2022DeepGANnel}.

Synthesis can also support restricted-data workflows. When real physiological datasets are difficult to share because of privacy or governance constraints, synthetic signals may provide shareable surrogates for preliminary method development or benchmarking \cite{ECGAN2023,Lange2024WearablePrivacy}. In wearable sensing, privacy-aware GAN-based synthesis has been used to generate multi-sensor smartwatch data for stress detection under differential privacy constraints, making the utility--privacy trade-off part of the generative objective \cite{Lange2024WearablePrivacy}. These uses show that synthesis is not limited to visually realistic waveform generation. It can support controlled experimentation, data sharing, and model development when real data are scarce, sensitive, or difficult to release.

The evaluation of synthesis depends strongly on intended use. Class-conditioned ECG generation requires label consistency and clinically meaningful morphology; fetal heart-rate synthesis requires category-specific temporal structure; glucose-trajectory synthesis requires subject-specific dynamics, event timing, and plausible physiological ranges; and privacy-preserving wearable synthesis requires a balance between downstream utility and memorization or leakage. More broadly, physiological-signal synthesis should move beyond unconstrained waveform sampling toward physiologically constrained and controllable generation. Synthetic signals should not only appear plausible, but also respect the physiological, measurement, and task-specific constraints implied by their conditions or intended use. Evaluation should therefore extend from distributional realism to physiological plausibility, condition consistency, diversity, and trustworthiness, as discussed in Section~VI.

\section{Model Families Across Roles and Modalities}
\begin{figure*}[!t]
\centering
\includegraphics[width=\textwidth]{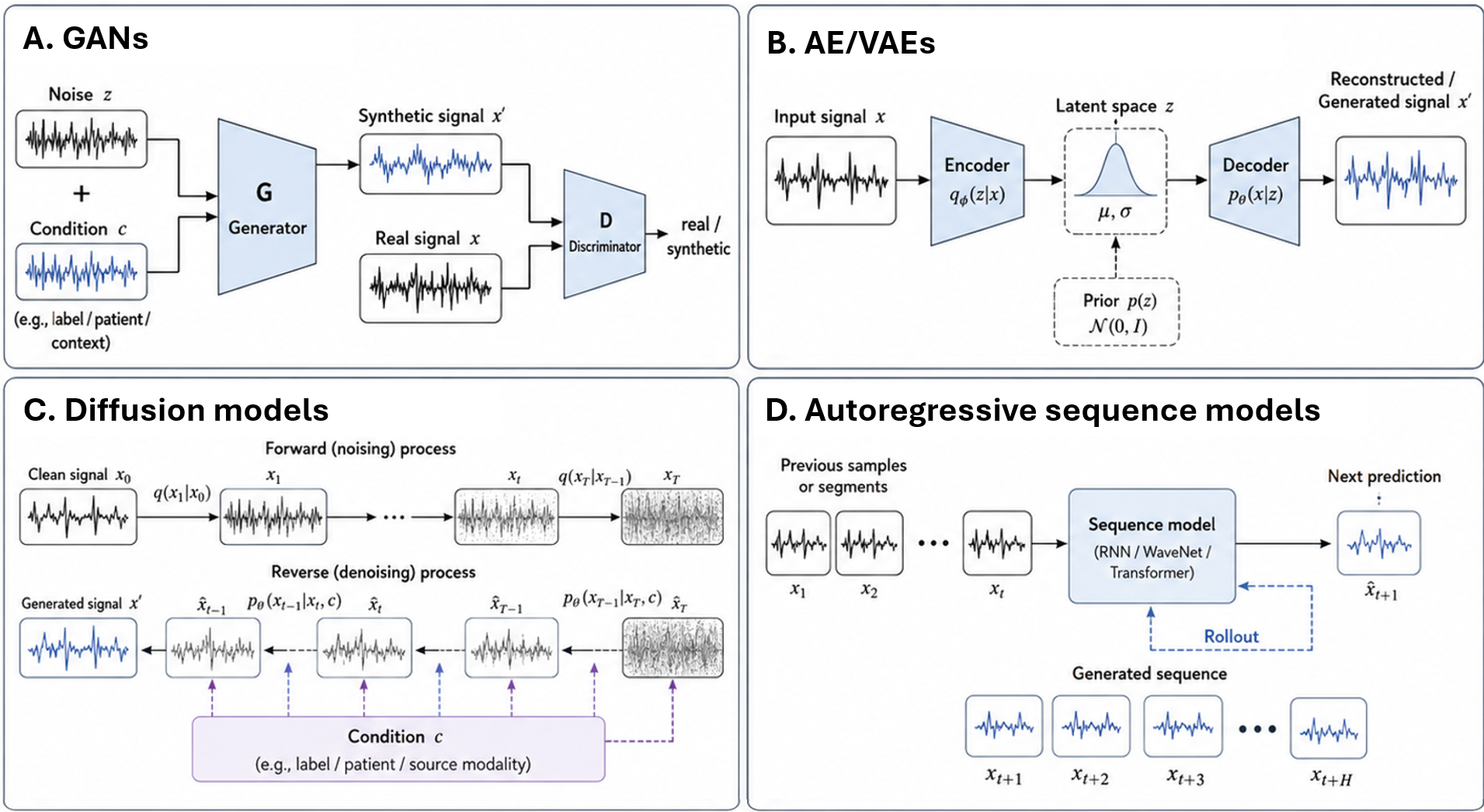}
\caption{Simplified workflows of major generative model families used in physiological-signal research. A. GAN-based models generate synthetic signals through adversarial training between a generator and discriminator. B. Autoencoding and variational models encode signals into latent representations and decode them for reconstruction, sampling, or representation learning. C. Diffusion and score-based models generate signals by learning an iterative denoising process. D. Autoregressive and sequence-based models generate future samples or segments from previous signal history.
}
\label{fig:models}
\end{figure*}

The reviewed studies can be broadly grouped into five recurring model-design patterns: GANs, autoencoding and variational models, diffusion models, autoregressive or sequence models, and hybrid models. These patterns differ in both architecture and generative principle. Figure~\ref{fig:models} summarizes the simplified workflows of the four core model families, while hybrid designs are discussed as task-driven combinations of these mechanisms. In practice, the same model family can support different functional roles. Table~\ref{tab:model_role_modality} maps representative studies across model family, functional role, and physiological signal family.

\begin{table*}[!t]
\centering
\caption{Representative generative physiological-signal studies mapped by functional role, model family, and signal family. Asterisks indicate studies that explicitly involved more than one functional role and were therefore placed in an overlapping or secondary role. Empty cells indicate that no representative study was selected for that combination in the analytical corpus, not that the combination is absent from the broader literature. AE/VAE denotes autoencoder- or variational-autoencoder-based methods; Autoregr./seq. denotes autoregressive and other sequence-based methods.}
\label{tab:model_role_modality}
\tiny
\renewcommand{\arraystretch}{1.05}
\resizebox{\textwidth}{!}{%
\begin{tabular}{llccccc}
\hline
Role & Model family & Cardiovascular & Neural & Muscular & Peripheral & Others \\
\hline
\multirow{5}{*}{Augmentation}
& GAN
& \cite{D0263}; \cite{D0168}; \cite{D0307}; \cite{D0139}; \cite{D0158}; \cite{TCGAN2023}; \cite{Kuntalp2024ClusterGAN} ; \cite{Qu2023QuantumECG} 
& \cite{D0133}; \cite{D0018}; \cite{D0187}; \cite{D0208}
& \cite{D0299}; \cite{D0298}; \cite{Chen2022EMG} 
& \cite{Jayalakshmy2021RespiratoryCGAN}; \cite{Ehrhart2022WearableStressCGAN} ; \cite{Furdui2021ACWGAN} 
& \cite{Seo2024CGMAugmentation}  \\
& AE/VAE
& \cite{D0141}\textsuperscript{*}
& \cite{D0180} 
& --
& \cite{D0020} \textsuperscript{*}; \cite{Saldanha2022RespiratoryVAE} 
& -- \\
& Diffusion
& \cite{Kang2024PPGDiffusion} 
& \cite{Torma2025EEGDiffusion}; \cite{D0163}; \cite{Chen2025EEGFNIRS} 
& --
& --
& -- \\
& Autoregr./seq.
& \cite{D0024}
& \cite{D0188}
& --
& --
& -- \\
& Hybrid/other
& --
& \cite{D0300}
& --
& --
& -- \\
\hline
\multirow{5}{*}{Restoration}
& GAN
& \cite{D0083}; \cite{D0032}; \cite{D0105}; \cite{D0135}; \cite{D0269}; \cite{Kiranyaz2022BlindECGRestoration} ; \cite{Zargari2023PPGCycleGAN} 
& \cite{D0301}; \cite{D0092}
& --
& --
& -- \\
& AE/VAE
& \cite{D0125}; \cite{D0017}; \cite{STMEM2024} 
& \cite{D0165}; \cite{MAEEG2022} 
& \cite{D0331} 
& \cite{Lin2020WearableMissing} 
& \cite{Ribeiro2023VagusNerveML} \\
& Diffusion
& \cite{D0152}; \cite{D0320}; \cite{D0333}; \cite{D0160}
& \cite{D0166}; \cite{D0328}
& \cite{D0057}; \cite{D0268}
& --
& -- \\
& Autoregr./seq.
& \cite{D0070}; \cite{D0184}
& \cite{D0058}; \cite{D0233}
& \cite{D0130}; \cite{D0297}
& --
& -- \\
& Hybrid/other
& --
& --
& --
& --
& -- \\
\hline
\multirow{5}{*}{Cross-modal}
& GAN
& \cite{D0059}; \cite{D0335}; \cite{CardioGAN2021} 
& \cite{D0048}; \cite{D0203}; \cite{D0258}; \cite{D0246}
& --
& \cite{Aqajari2021PPGRespirationCycleGAN} 
& -- \\
& AE/VAE
& --
& --
& --
& --
& -- \\
& Diffusion
& \cite{D0122}; \cite{RDDM2024} 
& --
& --
& --
& -- \\
& Autoregr./seq.
& \cite{Ding2025PPGECG}
& --
& --
& --
& -- \\
& Hybrid/other
& \cite{BioCross2025} 
& --
& --
& --
& -- \\
\hline
\multirow{5}{*}{Synthesis}
& GAN
& \cite{D0318}; \cite{FHRGAN2022}; \cite{D0002}; \cite{D0215}; \cite{D0319}; \cite{ECGAN2023} ; \cite{Neifar2024ShapePriors} 
& \cite{D0100}; \cite{D0053}; \cite{D0205}; \cite{D0327}; \cite{Li2025EEGPlausibility} 
& \cite{D0060}; \cite{D0076}
& \cite{Lange2024WearablePrivacy} 
& \cite{Cichosz2022CGMCGAN} ; \cite{Zhu2023GluGAN} ; \cite{Mujahid2022GlucoseCGAN} ; \cite{Ball2022DeepGANnel}  \\
& AE/VAE
& \cite{D0211}
& \cite{D0185}; \cite{D0061}; \cite{EEG2Vec2022} ; \cite{VAEEG2024} 
& --
& \cite{D0020}
& -- \\
& Diffusion
& \cite{SSSDECG2023}; \cite{D0302}; \cite{D0316}; \cite{D0127}; \cite{TransDiffECG2025} ; \cite{DiffuSETS2025} 
& \cite{D0073}; \cite{D0330}
& \cite{D0257}
& \cite{Choi2025EOGDiffusion} 
& -- \\
& Autoregr./seq.
& \cite{Wulan2020WaveNetECG} 
& --
& \cite{D0194}
& --
& -- \\
& Hybrid/other
& --
& \cite{TransCVAEGAN2025} 
& \cite{ChatEMG2025}
& --
& -- \\
\hline
\end{tabular}%
}
\end{table*}

\subsection{GANs}

As illustrated in Figure~\ref{fig:models}A, GANs generate synthetic signals through adversarial training between a generator and a discriminator. The generator $G$ maps latent noise $z$, labels, or other conditioning information to candidate signals, while the discriminator $D$ estimates whether an input signal is real or generated. In the standard GAN formulation, $G$ aims to produce synthetic samples $G(z)$ that match the empirical data distribution, whereas $D$ aims to distinguish real recordings $x\sim p_d$ from generated samples $G(z)$, with $z\sim p_z$. This adversarial game is commonly written as
\begin{equation}
\begin{aligned}
\min_G \max_D V(D,G)
=&\ \underset{x\sim p_d(x)}{\mathbb{E}}[\log D(x)] \\
&+\underset{z\sim p_z(z)}{\mathbb{E}}[\log(1-D(G(z)))] .
\end{aligned}
\end{equation}
where $p_d$ denotes the empirical distribution of real recordings, $p_z$ denotes the latent noise distribution, and $\mathbb{E}$ denotes expectation. Physiological-signal applications often extend this adversarial framework with conditioning variables, reconstruction losses, or task-specific constraints so that generated signals preserve labels, source modalities, subject context, or physiologically relevant structure. Across the reviewed literature, GAN-based models appear in all major functional roles, including augmentation, restoration, cross-modal generation, and synthesis \cite{D0263,D0059,Kiranyaz2022BlindECGRestoration,Zargari2023PPGCycleGAN}.

This pattern is especially clear in ECG and related cardiovascular applications, where GANs have been used to generate abnormal or minority-class beats, synthesize ECG or fetal heart-rate waveforms, and support classification-oriented augmentation \cite{TCGAN2023,ECGAN2023,FHRGAN2022}. Adversarial approaches have also been used for PPG-to-ECG translation, EMG augmentation, EEG generation, and multimodal wearable signal generation \cite{CardioGAN2021,Chen2022EMG,Furdui2021ACWGAN,Li2025EEGPlausibility}. The main risk is specific to adversarial learning: a discriminator may reward signals that look plausible within a dataset without ensuring that the generated waveform preserves clinically or physiologically meaningful structure. Mode collapse, training instability, discriminator overfitting, and incomplete minority-class coverage are particularly relevant when physiological datasets are small or imbalanced \cite{Goodfellow2014GAN}. For example, a generated ECG may retain class-discriminative features useful for a classifier while distorting interval structure, abnormal morphology, or multi-lead consistency. GAN-based studies are therefore strongest when adversarial realism is supported by explicit checks for diversity, condition consistency, and physiological plausibility.

\subsection{Autoencoding, variational, and reconstruction-based models}

As illustrated in Figure~\ref{fig:models}B, autoencoding models encode physiological signals into latent representations and decode them for reconstruction or representation learning. A plain autoencoder does not by itself define a prior distribution for principled sampling. A VAE, by contrast, learns an approximate posterior $q_{\phi}(z \mid x)$ over latent variables and regularizes it toward a prior $p(z)$, while the decoder $p_{\theta}(x \mid z)$ reconstructs the input signal \cite{Kingma2014VAE}. A common weighted VAE loss can be written as
\begin{equation}
\begin{aligned}
\mathcal{L}*{\mathrm{VAE}}
=&\ \mathbb{E}*{q_{\phi}(z \mid x)}
\left[-\log p_{\theta}(x \mid z)\right] \\
&+\beta,D_{\mathrm{KL}}
\left(q_{\phi}(z \mid x),|,p(z)\right),
\end{aligned}
\end{equation}
where $q_{\phi}(z \mid x)$ denotes the approximate posterior, $p(z)$ denotes the latent prior, $p_{\theta}(x \mid z)$ denotes the decoder likelihood, $D_{\mathrm{KL}}$ denotes the Kullback--Leibler divergence, and $\beta$ controls the relative weight of the reconstruction and regularization terms. For physiological signals, reconstruction should not be interpreted only as pointwise waveform error; depending on the modality, it may also need to preserve morphology, spectral content, channel relationships, event timing, or subject-specific dynamics. In this review, autoencoding and variational methods are especially relevant for latent representation learning, reconstruction, completion, and masked-signal learning from missing, corrupted, or unlabeled physiological recordings \cite{Lin2020WearableMissing,EEG2Vec2022,MAEEG2022,VAEEG2024,STMEM2024}.

Representative applications include masked EEG or ECG reconstruction, VAE-based neural-signal generation, and wearable missing-value recovery \cite{EEG2Vec2022,MAEEG2022,VAEEG2024,STMEM2024,Lin2020WearableMissing}. The strength of this model family lies in its encoding--decoding structure, which allows missing, compressed, or corrupted signal content to be inferred from available observations. Its main limitation is conservative reconstruction: when multiple physiological patterns are compatible with the same incomplete or noisy input, the decoder may favor a typical training-set pattern over a rare but clinically or functionally relevant event. Therefore, decoded samples should not be treated as useful augmentation or synthesis unless they are shown to recover or introduce meaningful physiological variation.

\subsection{Diffusion models}

As illustrated in Figure~\ref{fig:models}C, diffusion models generate signals through an iterative denoising process. Starting from a clean signal $x_0$, the forward process gradually adds Gaussian noise to produce noisier versions $x_1,\dots,x_T$, while a neural network is trained to reverse this process step by step. A common forward transition can be written as
\begin{equation}
q(x_t \mid x_{t-1}) =
\mathcal{N}\left(x_t;\sqrt{1-\beta_t},x_{t-1},\beta_t I\right),
\end{equation}
where $\mathcal{N}$ denotes a Gaussian distribution, $\sqrt{1-\beta_t},x_{t-1}$ is the mean, and $\beta_t I$ is the covariance determined by the noise schedule $\beta_t$. In the standard noise-prediction formulation, the model learns to predict the injected noise $\epsilon$ from a noisy sample $x_t$, the diffusion step $t$, and optional conditioning information $c$:
\begin{equation}
\mathcal{L}_{\mathrm{diff}}
=
\mathbb{E}_{t,x_0,\epsilon}
\left[
\left\|
\epsilon-\epsilon_{\theta}(x_t,t,c)
\right\|_2^2
\right].
\end{equation}
Here, $\epsilon_{\theta}$ denotes the learned denoising network, $x_t$ is the noisy signal at diffusion step \textit{t}, $\epsilon$ is the injected noise, and \textit{c} denotes optional conditioning information such as a class label, source modality, patient metadata, or physiological attribute \cite{Ho2020DDPM}.

Recent ECG studies illustrate this direction. SSSD-ECG used diffusion for multi-label conditional 12-lead ECG generation, TransDiffECG introduced semantically controllable ECG synthesis, and DiffuSETS extended ECG generation to text- and patient-conditioned settings \cite{SSSDECG2023,TransDiffECG2025,DiffuSETS2025}. Diffusion-based methods have also been applied to PPG generation, PPG-to-ECG translation, EEG augmentation, and EEG--fNIRS applications \cite{Kang2024PPGDiffusion,RDDM2024,Torma2025EEGDiffusion,Chen2025EEGFNIRS}. Compared with GANs, diffusion models are attractive because generation is based on learned denoising rather than adversarial discrimination. For physiological signals, this advantage is meaningful only when small but important structures are preserved, such as ECG interval changes, EEG event-related components, EMG burst timing, or PCG landmarks. The main practical limitation is the computational cost of iterative sampling \cite{DDIM2021,Luo2025BiomedicalDiffusion}. In conditional applications, generated signals should also be tested for whether they reflect the specified condition rather than merely appearing plausible.

\subsection{Autoregressive sequence models}

As illustrated in Figure~\ref{fig:models}D, autoregressive sequence models generate a signal by predicting each future sample or segment from previous signal history. For a physiological sequence $x_{1:T}$ and optional conditioning information $c$, this can be expressed as
\begin{equation}
p(x_{1:T}\mid c)=\prod_{t=1}^{T}p(x_t\mid x_{<t},c),
\end{equation}
where $x_{<t}$ denotes the previous signal context. This formulation is most relevant when temporal continuity, sample-level dependence, or rollout behavior is central to the task. A WaveNet-based ECG generator, for example, produces continuous ECG by modeling sample-level temporal structure \cite{Wulan2020WaveNetECG}. Recurrent, convolutional, or transformer-based sequence models may also be used when longer temporal context or sequence-level control is required.

The main limitation is error accumulation during rollout: locally plausible predictions may gradually drift away from long-range physiological consistency. Sequence models should therefore be evaluated not only by local waveform similarity, but also by temporal coherence, rhythm stability, event timing, and long-horizon consistency.

\subsection{Hybrid and task-aware models}

Hybrid models are not a single generative paradigm. They combine generative mechanisms, network architectures, and task-specific objectives to meet the requirements of a given physiological-signal problem. These combinations become useful when generation must preserve waveform morphology, use clinical or subject-level conditions, align multiple modalities, incorporate physiological priors, or produce annotations together with the signal.

Representative examples include Trans-cVAE-GAN for EEG generation, BioCross for cross-modal alignment among ECG, PPG, and arterial blood pressure, ECGAN and statistical-shape-prior GANs for rhythm-aware or prior-informed ECG synthesis, and DeepGANnel for patch-clamp trace generation with aligned annotations \cite{TransCVAEGAN2025,BioCross2025,ECGAN2023,Neifar2024ShapePriors,Ball2022DeepGANnel}. The main risk of hybrid models is attribution. When several components are introduced at once, it becomes difficult to determine which part improves realism, diversity, physiological plausibility, or downstream utility. Hybrid models are therefore most convincing when ablation studies and matched baselines show that the added mechanisms are necessary for the stated physiological task.

Overall, model choice in physiological-signal generation should be driven by role, signal family, and evaluation target rather than architectural novelty alone. GANs emphasize adversarial realism, autoencoding and variational models emphasize latent reconstruction or regularized sampling, diffusion models emphasize learned denoising and conditional sampling, autoregressive models emphasize temporal rollout, and hybrid models combine mechanisms for task-specific constraints. The evidence summarized in Table~\ref{tab:model_role_modality} supports a practical conclusion: generative physiological-signal modeling is not a single technical problem, and model families should be compared within the functional and physiological settings in which they are used.

\section{Evaluation Practices and Evidence Levels}

Evaluation remains inconsistent across generative physiological signal studies. Many studies rely on a single form of evidence, such as waveform similarity, feature-space overlap, or downstream classifier improvement \cite{Berger2023,Wang2025ECGEval,Zanchi2025SyntheticECG}. Although each is informative, they support different claims. Low reconstruction error does not establish physiological validity; similarity to the observed dataset does not demonstrate coverage of rare or unobserved variation; and improved classification may reflect task-relevant features without showing that the generated signal is physiologically complete.

\begin{figure}[!t]
\centering
\includegraphics[width=\columnwidth]{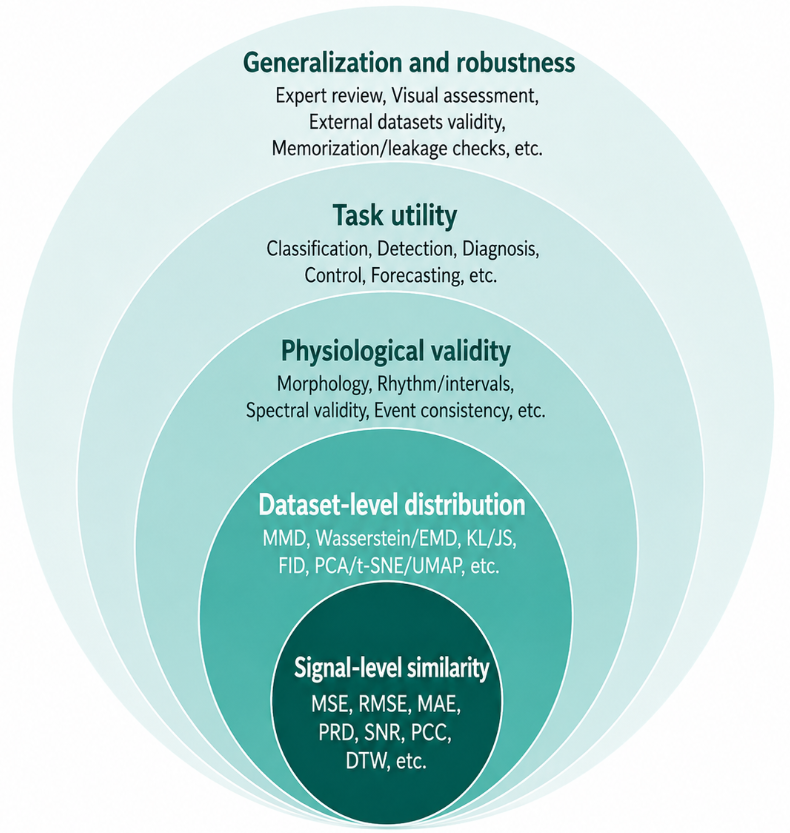}
\caption{
Layered evaluation targets for generative physiological signals.
Signal-level similarity evaluates pointwise or waveform-level agreement with reference signals.
Dataset-level distribution evaluates whether generated samples occupy a similar statistical space to real recordings.
Physiological validity evaluates whether generated signals preserve modality-relevant biological structure.
Task utility evaluates whether generated signals support the intended downstream use.
Generalization and robustness evaluate whether evidence holds beyond the original training setting and whether memorization or leakage is unlikely.
}
\label{fig:evaluation}
\end{figure}

This review therefore organizes evaluation as a claim-dependent evidence hierarchy. Signal-level similarity and dataset-level distribution assess resemblance to reference signals or available data. Physiological validity examines whether signal-specific biological structures are preserved, while task utility evaluates whether generated signals support the intended downstream use. Generalization and robustness assess whether these findings remain credible beyond the original dataset or recording setting and are not driven by leakage or memorization. Figure~\ref{fig:evaluation} defines the five evidence levels, and Table~\ref{tab:role_evaluation} maps them to role-specific evaluation priorities and common risks. The framework is not a mandatory checklist or a fixed ranking of importance; the evidence required depends on the claim being made.

\subsection{Signal-level similarity}

Signal-level similarity is the most direct form of evaluation. Generated, reconstructed, translated, or denoised signals are compared with real or reference signals using metrics such as mean squared error (MSE), root mean squared error (RMSE), mean absolute error (MAE), percentage root-mean-square difference (PRD), signal-to-noise ratio (SNR), Pearson correlation coefficient (PCC), or dynamic time warping (DTW) \cite{Bing2024ECGDenoising,Tang2022ECGReconstruction,Berndt1994DTW}. These metrics are especially common when a paired reference exists, as in restoration, denoising, imputation, or source-to-target translation.

This level is useful because a signal that fails basic waveform, temporal, or spectral agreement is unlikely to be credible. In cross-signal translation, for example, PPG-to-ECG studies use multiple similarity metrics to assess whether the reconstructed ECG resembles the target ECG waveform \cite{RDDM2024,Ding2025PPGECG}. In denoising or restoration, error and SNR-based measures can quantify whether a corrupted signal has been brought closer to a cleaner reference. However, signal-level similarity is a local form of evidence. A low error does not guarantee that clinically important morphology, rare events, phase relationships, or abnormal patterns have been preserved. This is why ECG evaluation-focused reviews argue that morphology and functional consistency should be assessed together rather than relying on any single signal-level metric \cite{Wang2025ECGEval}.

\subsection{Dataset-level distribution}

Dataset-level distribution evaluates generated samples as a set rather than as individual paired signals. It is especially relevant for synthesis and augmentation, where generated samples are often not paired with a specific reference recording. Common approaches include maximum mean discrepancy (MMD), Wasserstein or earth mover's distance (EMD), Kullback--Leibler (KL) or Jensen--Shannon (JS) divergence, Fr'echet inception distance (FID)-like feature distances, and feature-space visualization using principal component analysis (PCA), t-SNE, or Uniform Manifold Approximation and Projection (UMAP) \cite{Neifar2025}. These methods assess whether generated and real signals occupy similar statistical or feature-space distributions. FID-like scores require particular care because the original FID was defined for image features; in physiological-signal studies, their validity depends on whether the feature extractor captures task- and signal-relevant structure.

Distributional evaluation is useful because a generator may produce individually plausible samples while failing to represent the broader dataset. It can reveal whether synthetic samples cover the real feature space, collapse into narrow regions, or remain separable from real samples in handcrafted or learned representations. However, this level of evaluation still mainly tests resemblance to the available data. It does not show whether generated samples cover missing real-world variation, adequately represent rare physiological states, or avoid memorizing training examples. For augmentation, this distinction is critical: matching the training distribution is not the same as expanding useful variation for a downstream task.

\subsection{Physiological validity}

Physiological validity evaluates whether generated signals preserve biologically interpretable structures rather than only generic time-series properties. These structures are modality-specific: ECG validity may depend on P-QRS-T morphology, rhythm, interval structure, abnormal patterns, and multi-lead consistency; EEG validity may depend on event-related potentials, spectral organization, bandpower, entropy, and state-dependent dynamics \cite{Li2025EEGPlausibility}; EMG validity may depend on burst timing and movement-related activation; and fetal heart rate, PCG, ABP, and other specialized traces may require longer temporal patterns or event landmarks, such as accelerations, decelerations, S1/S2 timing, systolic and diastolic features, dicrotic points, or aligned trace annotations \cite{Homaeinezhad2012,Ball2022DeepGANnel}.

This level is less standardized than signal-level or distribution-level evaluation. Generic waveform metrics can be weak proxies: a generated ECG may have low RMSE while attenuating subtle ST-segment elevation or altering interval structure, and an EEG trace may appear realistic while distorting ERP morphology, spectra, or bandpower \cite{Li2025EEGPlausibility}. Physiological considerations should therefore inform not only post hoc validation, but also metric and model design. For example, in \cite{bao2026diffusion}, rhythm score, explosion score, and best-peak lag were used as exploratory metrics to constrain generated PCG. Rhythm-conditioned and shape-prior ECG synthesis provides a similar example, where physiological constraints are incorporated into the generation or evaluation process \cite{ECGAN2023,Neifar2024ShapePriors}.

\subsection{Task utility}

Task utility evaluates generated signals through their effect on downstream models, most commonly in classification, detection, recognition, forecasting, or control tasks. It is a task-based rather than signal-intrinsic evaluation: generated data are useful if they improve performance for the intended task under a valid real-data test protocol. Many studies assess this by training models with real and generated data and reporting downstream classification performance \cite{Neifar2025}. Examples include ECG augmentation for minority-class recognition, synthetic sleep EEG for few-shot sleep staging, and generated EMG for movement classification or intent inference in robotic hand-orthosis control \cite{Adib2025SyntheticECG,Tang2025CCGAN,TCGAN2023,SleepGAN2022,Coelho2023,Chen2022EMG,ChatEMG2025}.

Task utility should be interpreted within this purpose. For augmentation, generated samples should be used only in training or data augmentation, and performance should be compared with a real-data baseline on held-out real recordings. Otherwise, reported gains may reflect leakage, subject or session overlap, or recording-specific shortcuts. For restoration and cross-modal generation, utility requires that the denoised, completed, or translated signal retain information needed by the downstream model \cite{RDDM2024,Ding2025PPGECG}. When the goal of generation is mainly to improve recognition or prediction, this task-based evidence may be sufficient for that claim. However, it should not be taken as proof of full physiological validity or transferability beyond the tested setting.

\subsection{Generalization and robustness}

Generalization and robustness evaluate whether evidence from the previous levels remains credible beyond the original dataset or data split. This level includes external validation, subject-independent testing, robustness to acquisition differences, and checks for leakage or memorization. These checks are important because physiological signals are strongly shaped by subject variability, sensor placement, device characteristics, and recording context. A generator that appears effective within one dataset may therefore fail under a different recording condition \cite{ChatEMG2025,Li2025WearableAffective,VanDerDonckt2024,Jafari2025Transfer}.

The required evidence depends on the claim. For a narrow task-oriented claim, such as improving a classifier through augmentation, a strict held-out real test set may be sufficient. For broader claims, such as reusable synthesis, data sharing, or clinical-facing use, stronger evidence is needed. Generated signals should remain physiologically plausible, avoid reproducing training examples, and retain utility under external or shifted conditions. Expert assessment, external utility testing, and privacy-aware evaluation can help identify failures missed by standard metrics \cite{bao2026diffusion,Russo2025ECGEval,Lange2024WearablePrivacy,Senthuran2025ECGPrivacy}.

\begin{table}[!t]
\centering
\caption{Role-specific evaluation priorities and common risks.}
\label{tab:role_evaluation}
\scriptsize
\setlength{\tabcolsep}{2pt}
\renewcommand{\arraystretch}{1.12}

\begin{tabular}{@{}
>{\raggedright\arraybackslash}p{0.16\columnwidth}
>{\raggedright\arraybackslash}p{0.45\columnwidth}
>{\raggedright\arraybackslash}p{0.27\columnwidth}
@{}}
\hline
\textbf{Role} &
\textbf{Priority evidence} &
\textbf{Main risk} \\
\hline

Augmentation &
Improvement over real-data-only training on held-out real data; independent splits. &
Leakage, near-duplicates, or shortcuts. \\

Restoration &
Reference agreement and preservation of morphology, events, or spectra. &
Over-smoothing or loss of abnormalities. \\

Cross-modal &
Target fidelity, source--target consistency, and uncertainty for ambiguous mappings. &
Unsupported target features. \\

Synthesis &
Condition fidelity, physiological validity, diversity, and external validation. &
Mode collapse or overclaimed realism. \\

\hline
\end{tabular}
\end{table}

Taken together, the five evidence levels and the role-specific priorities in Table~\ref{tab:role_evaluation} show that evaluation should be matched to the intended claim rather than reduced to a universal metric set. No single level is sufficient for every use: evidence of resemblance, physiological validity, task utility, and generalization addresses different aspects of the generated signal. This claim-dependent view provides the basis for the following discussion of physiological synthesis, model design, transparent reporting, and responsible use.

\section{Discussion and Future Directions}

The reviewed literature shows that generative modeling has become a broad methodological tool for physiological-signal research. Its current use, however, remains largely functional. Most studies use generation to augment scarce data, restore corrupted recordings, translate between modalities, protect data sharing, or improve downstream models. In these settings, the generated signal is usually not the final scientific object, but an intermediate tool used to support another task.

This distinction is important. Unlike image or speech generation, where the generated output is often the intended product, physiological-signal generation is usually judged by what the generated signal enables: better classification, more robust training, signal completion, modality translation, or controlled analysis. The central question is therefore not simply whether a generated waveform appears realistic, but whether it preserves the information required for its intended use. This claim-dependent view also explains why evaluation practices differ across the field and why no single metric can be sufficient for all generative physiological-signal studies.

\subsection{Functional generation versus physiological synthesis}

A central distinction in the reviewed literature is between functional generation and physiological synthesis. In many studies, generative models are not used to produce standalone physiological signals as the final scientific object. They are used to support another task, such as data augmentation, restoration, cross-modal translation, privacy-oriented data sharing, or downstream model improvement \cite{Neifar2025,He2025TimeVarying,Loni2025Review}. In these settings, the generated signal is judged mainly by whether it preserves the information required for that function.

Physiological synthesis makes a stronger claim. Here, the generated signal is expected to be meaningful as a physiological object, not only as a useful input to another model. Such a claim requires evidence that signal-specific structures, clinical or physiological conditions, and source--target relationships are preserved. Treating functional generation and physiological synthesis as equivalent risks either overburdening task-oriented tools with unnecessary synthesis-level requirements or overclaiming signals that are useful only for a narrow downstream task.

\subsection{Toward controllable and physiology-aware generation}

A likely next step is to move from generic sample generation toward controllable and physiology-aware generation. Current functional uses remain valuable, but they are often constrained by the source data: a generator trained on weak labels, a narrow cohort, or a single recording protocol cannot be expected to create reliable physiological variation that is absent or poorly represented in the training set. More informative generation requires explicit conditions, such as rhythm type, disease label, patient context, acquisition setting, or source modality, and evidence that these conditions are reflected in the output. Recent ECG studies using physiological attributes, clinical text, or patient-specific information illustrate this direction \cite{TransDiffECG2025,DiffuSETS2025}.

Physiology-aware generation also requires model design to follow the structure of the target signal. The relevant constraints may involve morphology, rhythm, spectra, channel relationships, temporal landmarks, or cross-modal consistency. Rhythm-conditioned and shape-prior ECG synthesis shows how physiological assumptions can be incorporated into generation or evaluation \cite{ECGAN2023,Neifar2024ShapePriors}, while EEG--fNIRS and multimodal cardiovascular models illustrate how related signals can be modeled jointly or translated across modalities \cite{Chen2025EEGFNIRS,BioCross2025}. The methodological challenge is therefore not only to generate plausible waveforms, but to generate signals whose controlled attributes and physiological structures are testable.

\subsection{Claim-dependent evaluation and transparent reporting}

Evaluation should be matched to the claim being made. A study that uses generation only for augmentation does not need to prove full physiological synthesis, but it should show improved downstream performance under a valid real-data test protocol. A restoration study should show that meaningful signal content is preserved, not only that numerical error is reduced. A cross-modal generation study should assess both target-signal quality and source--target consistency. A study claiming reusable or controllable synthesis requires broader evidence, including signal-specific validity, condition consistency, and robustness beyond the original dataset \cite{Wang2025ECGEval,Russo2025ECGEval}.

Transparent reporting is therefore part of evaluation. Studies should state the intended use of generation, the signal structures or conditions expected to be preserved, the model inputs and conditioning variables, and the data split used for testing. When generated data are used for augmentation, reports should clarify whether test recordings, subjects, sessions, or devices are separated from the generation and training process. When synthetic data are proposed for sharing or privacy protection, utility should be reported together with memorization or leakage checks \cite{Lange2024WearablePrivacy,Senthuran2025ECGPrivacy}. These practices would not standardize all evaluation across signal types, but they would make claims easier to interpret and reduce overstatement based on narrow metrics.

\subsection{Boundaries and responsible use of synthetic physiological data}

Synthetic physiological data should be tied to a clearly stated purpose. Generating more samples does not automatically solve weak labels, biased cohorts, poor recording quality, missing clinical context, or unclear target outcomes. A data-driven generator can reproduce, interpolate, or reorganize patterns present in the source data, but it should not be assumed to create valid physiological diversity that was absent, incorrectly labeled, or confounded with recording conditions \cite{Neifar2025,Wang2025ECGEval}. In some settings, generation is an appropriate intervention; in others, better acquisition, stronger annotation, clearer endpoints, external validation, or more transparent benchmarks may matter more than a more complex generator.

Responsible use becomes especially important when synthetic physiological data are proposed for data sharing, privacy-preserving modeling, clinical-facing applications, or regulatory evidence. Existing medical-device and AI-enabled software frameworks do not make synthetic data unacceptable, but they require evidence that is traceable to the intended role of the data \cite{FDA2025AISaMD,EU2017MDR,EC2026MDCGGuidance}. A synthetic dataset that is useful for early benchmarking or task-specific augmentation is not automatically sufficient for clinical validation. Utility, leakage risk, privacy protection, bias, and preservation of health-relevant information should therefore be evaluated together when synthetic physiological data are used beyond exploratory modeling \cite{Kaabachi2025PrivacyUtility,Miletic2025LongitudinalSynthetic,Lange2024WearablePrivacy,Senthuran2025ECGPrivacy}.

\section{Conclusion}

Generative modeling has become an important methodological tool for physiological-signal research, extending beyond de novo waveform synthesis to augmentation, restoration, cross-modal generation, privacy-oriented data sharing, and controllable synthesis. This review organized the field across signal families, functional roles, model families, and evaluation evidence. The main conclusion is that generated physiological signals should be evaluated according to the claim they are intended to support. Similarity metrics and distributional comparisons are useful, but they cannot by themselves establish physiological validity, task utility, or generalizability. Future progress will depend on clearer task definitions, signal-specific constraints, transparent reporting, and stronger evaluation designs that distinguish functional usefulness from realistic and reusable physiological synthesis.

\appendices
\section{Detailed Search Strategy}
\label{app:search}

Searches were conducted in Web of Science, Scopus, PubMed, IEEE Xplore, and arXiv for records published from January 2015 to December 2025. Database-specific syntax was adapted where necessary, but the search logic followed the same four concept blocks: generative intent, generative model family, signal context, and physiological modality. Searches were restricted primarily to title and abstract fields, except in Scopus, where physiological-modality terms were also allowed in keywords. In IEEE Xplore, an abstract-only command search was used because of platform constraints on complex fielded queries.

The core search logic combined the following concepts: ('synthetic data' OR 'synthetic signal' OR 'synthetic waveform' OR 'signal generation' OR 'time series generation' OR 'data augmentation' OR imputation OR denoising OR inpainting) AND (GAN OR 'generative adversarial' OR VAE OR 'variational autoencoder' OR 'diffusion model' OR DDPM OR `score-based' OR `normalizing flow' OR autoregressive OR WaveNet) AND (signal OR waveform OR 'time series' OR biosignal OR recording OR sensor OR monitoring) AND (ECG OR EKG OR electrocardiogram OR EEG OR electroencephalography OR EMG OR electromyography OR PPG OR photoplethysmography OR PCG OR phonocardiogram OR 'heart sound' OR respiration OR 'blood pressure' OR SpO2 OR oximetry OR polysomnography OR 'sleep staging' OR cardiotocography OR 'fetal heart rate' OR electrodermal OR 'galvanic skin response'). Records related to systematic reviews, meta-analyses, evidence synthesis, or non-generative uses of 'data synthesis' were excluded where possible.

The initial search identified 379 records. After aggregation across databases, records were screened using titles, abstracts, and available full texts. Records were excluded if they did not involve physiological waveform or time-series data, lacked a substantive generative-modeling component, focused only on imaging or text, or mentioned synthetic data only peripherally. This process retained 256 studies as the content-eligible corpus. A representative subset of 120 studies was selected for in-depth analysis. The selection was designed to preserve coverage across major signal families, generative roles, and model families while reducing redundancy among studies addressing similar problems. The full content-eligible corpus was retained for corpus-level mapping and trend analysis. The corpus supported a structured review with corpus-based mapping and representative in-depth synthesis, rather than a formal systematic review or meta-analysis.

\bibliographystyle{IEEEtran}
\bibliography{IEEEabrv,Bibliography}
\end{document}